\documentclass[twocolumn]{jpsj2} 
\title{
Increase of Superconducting Correlation due to Dimensionality Change in Quasi-One-Dimensional Conductors
}

\author{Yuki \textsc{Fuseya}\thanks{E-mail address: fuseya@hosi.phys.s.u-tokyo.ac.jp}
and 
Masao \textsc{Ogata}}

\inst{Department of Physics, University of Tokyo,
Hongo, Bunkyo-ku, Tokyo 113-0033}

\abst{
	Response functions for spin-density-wave (SDW) and $d$-wave singlet superconductivity ($d$SC) in quasi-one-dimensional (Q1D) electron systems are calculated by a renormalization group technique.
	It is shown that the response functions for both SDW, $\chi_\rs$, and $d$SC, $\chi_d$, are enhanced by interchain hopping, $t_\perp$, i.e., by quasi-one dimensionality.
	When the Fermi surface deviates from perfect nesting, $\chi_\rs$  saturates below the energy scale of imperfectness of the nesting, while $\chi_d$ is hardly affected.
	Consequently, the superconducting correlation increases even away from the SDW phase.
	This gives a possible interpretation of the recent experimental results of Q1D organic conductor (TMTTF)$_2$SbF$_6$, where $T_{\rm c}$ increases even away from the SDW phase.
}

\kword{superconductivity, quasi-one-dimension, renormalization group, dimensionality, spin fluctuation, superconducting fluctuation, organic superconductor, (TMTTF)$_2$SbF$_6$}

\newcommand{\bk}{{\bf k}}
\newcommand{\bq}{{\bf q}}
\newcommand{\bQ}{{\bf Q}}

\newcommand{\rs}{{\rm s}}

\newcommand{\Tc}{T_{\rm c}}

\begin{document}
\maketitle

	%
	When a superconducting phase appears next to some ordered phase, it is reasonable to consider that the mechanism of the superconductivity is due to the fluctuation of the ordered phase.
	Since the amplitude of fluctuation is reduced as the system is away from the ordered phase, we expect that the transition temperature of superconductivity, $\Tc$, is also reduced {\it monotonically}.
	For example, in quasi-one-dimensional (Q1D) organic conductors TMTSF- and TMTTF-salts, its superconducting phase appears next to the spin-density-wave (SDW) phase, and $\Tc$ decreases monotonically as the system leaves away from the SDW phase by applying pressure\cite{Jerome}.
	This situation have been considered to be a typical example of the spin-fluctuation mediated superconductivity.\cite{Emery,Shimahara,KK,DB}
	The pressure-temperature ($P$-$T$) phase diagram in these salts have been interpreted by the simple spin-fluctuation mechanism\cite{Shimahara}.

	However, a recent careful measurement of (TMTTF)$_2$SbF$_6$ shows that $\Tc$ increases by pressure even away from the SDW phase\cite{Itoi}.
	This behavior of $\Tc$ cannot be understood from the above simple spin-fluctuation mechanism.
	It suggests a possibility that some other unknown effects enhance $\Tc$.

	Our idea here is to take account of the dimensionality change due to the pressure as a key to understand this mysterious behavior.
	So far, the effect of pressure has been considered to be just to destroy nesting properties of the Fermi surface and to reduce the spin-fluctuation.
	However, the pressure simultaneously increases the dimensionality through the increase of interchain hopping, $t_\perp$.
	The change of the dimensionality possibly affects the superconducting correlation.
	A pioneering work of this subject was done by Suzumura and Fukuyama\cite{SF}, who showed $\Tc$ of full gapped ($s$- or $p$-wave) pairing is an increasing function of $t_\perp$.
	Though their argument is applicable to the system with attractive interaction, it cannot be directly applied to the repulsive cases.
	There are some issues to be taken into account: interchain interaction, anisotropic pairing, such as $d$- or $f$-wave, and competition between superconductivity and SDW.
	These points have been difficult to consider theoretically so far.
	However, due to the improvement of the renormalization group (RG) technique\cite{DB,Fuseya2006}, we are now able to study these problems in a reliable way.
	
	%
	%
	%
	%
	%
	
	In this letter, we investigate spin- and superconducting-fluctuation in Q1D electron systems by changing the dimensionality and the nesting conditions with the newly developed RG technique for Q1D systems called $N$-chain RG\cite{Fuseya2006}.
	In particular, we clarify the following points: 1) is it possible for $\Tc$ to keep increasing even away from the SDW phase?, 2) where is the maximum of $\Tc$ as a function of pressure or of dimensionality?, 3) what are the roles of dimensionality and nesting property?
	In order to discuss the effect of dimensionality, it is of the prime importance to deal with 1D fluctuation adequately, which results from the quantum interference between Cooper (particle-particle) and Peierls (particle-hole) channels.
	Since this is difficult in RPA or FLEX type approximation\cite{Shimahara,KK}, we employ 
	the RG method, which has been approved for taking into account the 1D fluctuation\cite{Solyom,Bourbonnais}.
	%
	%
	The obtained results indicate that the superconducting correlation essentially increases for small $t_\perp$, and can keep increasing even when the spin fluctuation is suppressed.
	%

	
	We study the Hubbard model in weakly-coupled $N$ chains with length $L$.
	In particular, we investigate this model at quarter-filling, which models TMTSF- and TMTTF-salts.
	The present results, however, is valid for other fillings except for the half-filling.
	For the convenience of the RG formulation, we express the Hamiltonian in the form $H=H_0 + H_{\rm I}$ with
$
	H_0 = \sum_{p, \bk, \sigma} \xi_p (\bk ) c_{p, \bk, \sigma}^\dagger c_{p, \bk, \sigma}
$ and 
	%
\begin{align}
   H_{\rm I} &= \frac{1}{LN}\sum_{p, \bq}
   \bigl[
   g_{\rho}
   \rho_p (\bq) \rho_{-p} (-\bq)
   +g_{\sigma}
   {\bf S}_p (\bq)\cdot {\bf S}_{-p} (-\bq)
   \nonumber\\
   &+g_{4 \rho}
   \rho_p (\bq) \rho_p (-\bq)
   +g_{4\sigma}
   S_p^z (\bq)S_{p}^z (-\bq)
   \bigr] .
   \label{Hamiltonian}
\end{align}
	Here the coupling constants are given by 
	$g_\rho = g_{4\rho}=-g_\sigma=-g_{4\sigma}=U$ for the Hubbard model, $H_{\rm I}=U \sum_i n_{i\uparrow}n_{i\downarrow}$,
	and $c_{p, \bk, \sigma}$ ($c_{p, \bk, \sigma}^\dagger$) is an annihilation (creation) operator with spin $\sigma $ close to the Fermi surface of the right ($p=1$) and left ($p=-1$) branches.
	%
	%
	The charge and spin density operators of the branch $p$ are given by
$   \rho_p (\bq)
   	\equiv (1/2)\sum_{\bk, \sigma}
   	c_{p,\bk+\bq,\sigma}^\dagger
   	c_{p,\bk,\sigma}
   	$ and
%
   	${\bf S}_p (\bq)
   	\equiv (1/2)\sum_{\bk, \alpha, \beta}
   	c_{p,\bk+\bq,\alpha}^\dagger
   	\pmb{\sigma}^{\alpha \beta}
   	c_{p,\bk,\beta}.
$
	%
	With the Hamiltonian eq. (\ref{Hamiltonian}), $8k_F$-Umklapp process is discarded.
	In the low pressure region of TMTTF-salts, this Umklapp process is relevant and leads to the charge order transition.\cite{Yoshioka,Seo}
	In the high pressure region, on the other hand, the effect of the $8k_F$-Umklapp process is naively expected to be small.
	Thus we investigate the above Hamiltonian for the first step.
	%
	%
	
	We use the quasiparticle dispersion given by
	\begin{align}
		\xi_p (\bk) &= v_F (pk -k_F) 
		-2t_\perp \cos k_\perp -2t_{\perp2} \cos 2k_\perp,
		\label{dispersion}
	\end{align}
	where $v_F (= \sqrt{2}t)$ is the Fermi velocity at quarter-filling with $t$ being the intrachain hopping, $k_F$ being the Fermi wave number in the limit of $t_\perp=0$.
	$t_\perp$ and $t_{\perp2}$ are the nearest and the next-nearest interchain hopping, respectively.
	%
	%
	%
	%
	When $t_\perp>0$ and $t_{\perp2}=0$, the Fermi surface is perfectly nested with the nesting vector $\bQ_0 =  (\pm2k_F, \pm\pi)$ even though the Fermi surface is warped.
	On the other hand, when $t_\perp , t_{\perp2}>0$, the nesting becomes imperfect.
	These two parameters represent different effects.
	$t_\perp$ corresponds to the change of dimensionality, where the crossover from 1D to Q1D occurs by keeping the nesting properties, while $t_{\perp2}$ breaks the nesting condition.
	%
	%
	In the real compounds, we can change $t_\perp$ and $t_{\perp2}$ by applying pressure $P$.
	The relation between them depends on the various conditions of the experiments.

	We apply the $N$-chain RG to the above Hamiltonian.
	This technique has a strong point that it can deal with the 1D fluctuation effect even in the region $T\lesssim t_\perp$, where the system is not a non-Fermi liquid state within the one-loop approximation.
	The details of this RG method are described in ref. \citen{Fuseya2006}.
	The response functions are calculated from
\begin{eqnarray}
	\chi_{\mu } (T)=
	\frac{2}{\pi v_F }
	\int_0^\infty \!\! 
	d\ell \left[z_{\mu}(\ell)\right]^2
	I_{\rm CP}(\ell ),
\end{eqnarray}
	where $I_{\rm CP} = \tanh \left[ E/4T\right]$, $z_\mu$ is pair vertex part, and $\mu={\rm s}$ for SDW and $\mu=d$ for $d$SC.
	The initial band-width, $E_0 =2v_F k_F $, is renormalized as $E = E_0 e^{-\ell}$.
	The pair vertex part of the SDW is renormalized as 
\begin{eqnarray}
	\frac{d}{d\ell}\ln z_\rs (\ell) 
	=-\frac{1}{N^2}\sum_{k_\perp, k_\perp'}
	G_\rs (k_\perp, \pi -k_\perp' , k_\perp'; \ell)
	I_{\rm CP}(\ell).
\end{eqnarray}
	Here, $G_{\mu} (k_{1\perp}, k_{2\perp}, k_{3\perp}; \ell)$ is a renormalized four-point vertex at step $\ell$, where $k_{1\perp}$, $-k_{2\perp}$ and $k_{3\perp}$ are perpendicular components of incoming momenta $\bk_1, -\bk_2$, and outgoing momenta $\bk_3$, respectively.
	Note that $\bk_4$ is given by $-\bk_4 = \bk_3 + \bk_2 -\bk_1$.
	Generally, marginal interactions in the low-energy region depend only on polar coordinates (or in this case, $k_{i\perp}$) of the wave vectors.
	This is justified by the zero-order scaling analysis.\cite{Zanchi,Shankar}
	Thus the dependence of $G_\mu$ on momenta parallel to the chain is irrelevant.
	The initial value for the vertex is given by $G_{\mu} (\ell=0)\equiv g_\mu /\pi v_F$, where $g_\rs = (-g_\rho +g_\sigma)/4$ for SDW.
	For superconductivity, we consider partial components of the pair vertex.
	The pair vertex part of $d$SC is renormalized as
\begin{eqnarray}
	\frac{d}{d\ell}\ln z_d(\ell)
	=-a_dI_{\rm CP}(\ell),
\end{eqnarray}
	where
	$a_d = \left(4/N^2\right)
	\sum_{k_\perp, k_\perp'}\cos k_\perp 
	G_{\rm sin} (k_\perp, k_\perp, k_\perp') \cos k_\perp'$
	and
	$g_{\rm sin} = (g_\rho -3g_\sigma)/4$.
	(For the details of pairing symmetry, see ref. \citen{Fuseya2005b}.)
	%


\begin{figure}[tb]
\begin{center}
\includegraphics[width=7cm]{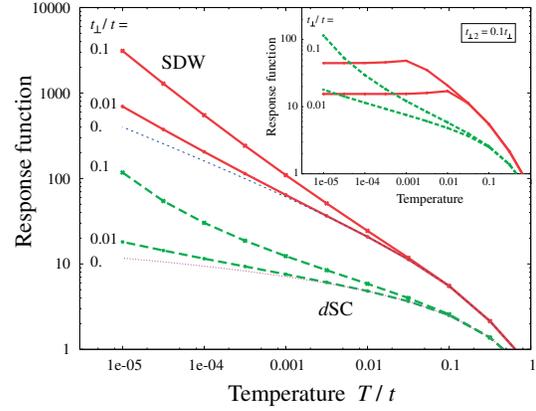}
\end{center}
\caption{Response functions of SDW (solid lines) and $d$SC (dashed lines) as a function of temperature for the perfectly nested Fermi surface $t_{\perp2}=0$ with $U=3t, N=15$, $t_\perp/t=0, 0.01, 0.1$.
Inset shows the response functions for the imperfect nesting case with $t_{\perp2}=0.1t_\perp$.}
\label{Fig1}
\end{figure}


	Figure \ref{Fig1} shows the temperature dependence of the response functions, $\chi_\mu (T)$, for $t_\perp /t= 0.0, 0.01, 0.1$ with $U=3t, N=15$.
	First we concentrate on the case where the Fermi surface is perfectly-nested, i.e., $t_{\perp2}=0$.
	For purely 1D systems ($t_\perp =0$), it is well known that $\chi_\rs (T)$ shows a power-law behavior as  $\chi_\rs (T) \propto T^{-\alpha_0}$ with $\alpha_0 = U/2\pi v_F$.
	In the case of Q1D systems ($t_\perp > 0$), $\chi_\rs (T)$ is enhanced from that of 1D, keeping apparently the power-law behavior as shown in Fig. \ref{Fig1}.
	The power $\alpha$ of Q1D $\chi_\rs(T)$ increases as $t_\perp$ increases and saturates for relatively large $t_\perp $ as shown in the inset of Fig. \ref{J}.
	%
	%
	%
	%
	The response function for $d$SC, $\chi_d (T)$, is also enhanced by $t_\perp$ and tends to diverge at low temperatures, though it does not show any instability at $t_\perp=0$ due to $a_d =0$.\cite{1dDelta}
	%
	%
\begin{figure}[tb]
\begin{center}
\includegraphics[width=7cm]{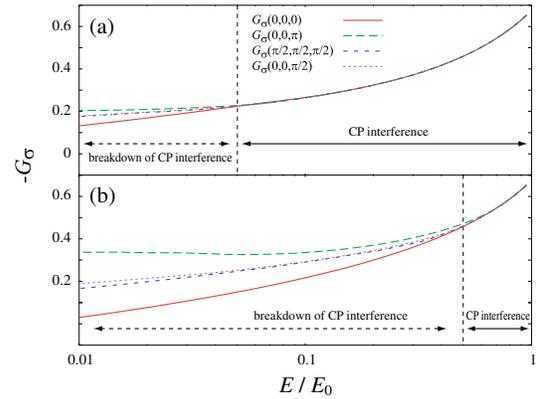}
\end{center}
\caption{Energy flows of the coupling $G_\sigma (k_{\perp1}, k_{\perp1}, k_{\perp3})$ for (a) $t_\perp =0.01t$ and (b) $t_\perp = 0.1t$ with $t_{\perp2}=0$, $U=3.0t$ at $T=10^{-6}t$.}
\label{Flow}
\end{figure}

	%
	
	The instability of $d$SC originates from the $k_\perp$-dependence of the coupilings.
	Figures \ref{Flow} show some examples of energy flow of $-G_\sigma (k_{1\perp}, k_{1\perp}, k_{3\perp})$ for several sets of $(k_{1\perp}, k_{3\perp})$, which corresponds to the Cooper pair scattering $(\bk_1, -\bk_1) \to (\bk_3, -\bk_3)$, with $t_{\perp2}=0$ and $U=3t$.
	In the high energy region $E\gtrsim E_{\rm x}\sim \mathcal{O}(t_\perp)$, every flow is identical with that of 1D.
	In this region, interference between Cooper (C) and Peierls (P) channels is dominant even though $t_\perp >0$, so that the couplings flow as those of 1D.
	For $E\lesssim E_{\rm x}$, on the other hand, the CP interference becomes imperfect, namely, C and P do not cancel each other in some types of couplings.\cite{Fuseya2006}
	As a result, each couplings flows differently and $G_\mu (k_{1\perp}, k_{2\perp}, k_{3\perp})$ has $k_\perp$-dependence.
	The amplitude of the $k_\perp$-dependence of the couplings becomes larger as $t_\perp$ becomes larger.
	This $k_\perp$-dependence causes the increase of $\chi_d$.
	In other words, the increase of $\chi_d$ is due to the breakdown of the perfect CP interference.
	This is a characteristic behavior of the Q1D systems.
	In two or three dimensions, there are no perfect CP intereference in high-energy region, and thus the increase of $\chi_d$ in the present context does not occur.
	Actually, for the isotropic square-lattice, the RG method yields the monotonically decrease of $\Tc$\cite{Zanchi,Halboth,Honerkamp}.
	%
	%

\begin{figure}[tb]
\begin{center}
\includegraphics[width=7cm]{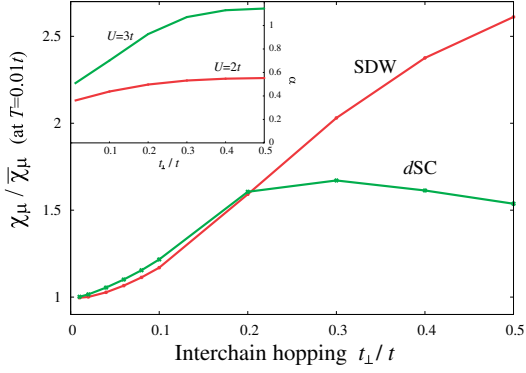}
\end{center}
\caption{Response function of Q1D normalized by that of 1D, $\chi_\mu / \bar{\chi}_\mu$, as a function of $t_\perp$ for $t_{\perp2}=0$, $U=3t$ and $T=0.01t$.
The inset shows the power of $\chi_\rs (T)$ $(\propto T^{-\alpha})$ for $U/t=2,3$.
}
\label{J}
\end{figure}
\begin{figure}[tb]
\begin{flushleft}
\hspace{33mm}
\footnotesize
(a) SDW \hspace{30mm} (b) $d$-wave SC
\end{flushleft}
\begin{center}
\includegraphics[width=7cm]{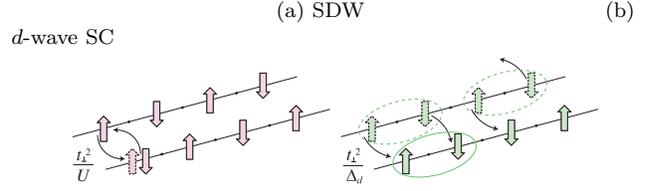}
\end{center}
\caption{Illustration of the interchain process for (a) SDW and (b) $d$SC. 
The long-range SDW order is stabilized by the interchain exchange coupling, which is the single-particle process. 
The long-range $d$SC order is stabilized by the interchain Josephson coupling, which is the pair process. The coherent pair hopping is suppressed for large $t_\perp$.}
\label{illust}
\end{figure}

	%
	%
	%
	Both $\chi_\rs$ and $\chi_d$ increase for small $t_\perp$ as we have seen, but there are difference for large $t_\perp$.
	We plot the response functions normalized by that of 1D, $\bar{\chi}_\mu$, as a function of $t_\perp$ for the perfect nesting case ($t_{\perp2}=0$) at $T=0.01t$ in Fig. \ref{J}.
	For small $t_\perp$ ($\lesssim 0.25t$), both $\chi_\mu / \bar{\chi}_\mu$ are proportional to $t_\perp^2$.
	%
	%
	For large $t_\perp$ ($\gtrsim 0.25t$), on the other hand, $\chi_\mu / \bar{\chi}_\mu$ behaves differently: $\chi_d / \bar{\chi}_d$ saturates and slightly decreases, whereas $\chi_\rs /\bar{\chi}_\rs$ keeps increasing.
	These different properties of $\chi_\mu /\bar{\chi}_\mu$ against $t_\perp$ can be understood intuitively as follows.
	For SDW, the exchange coupling between chains stabilizes the long-range SDW order\cite{Bourbonnais}, and enhances $\chi_\rs$. 
	%
	This process is a {\it single-particle} process as is shown in Fig. \ref{illust} (a), so that $J_\rs \sim t_\perp^2/U$.
	Here, $J_\mu$ is defined as $J_\mu \equiv \left( \chi_\mu / \bar{\chi}_\mu -1\right)/\bar{\chi}_\mu$, which can be regarded as an effective interchain-coupling.
	For $d$SC, on the other hand, the interchain Josephson coupling stabilizes the $d$SC order and enhances $\chi_d$. 
	%
	This process is a {\it pair} process as is shown in Fig. \ref{illust} (b).
	%
	%
	In this case, $J_d \sim t_\perp^2/\Delta_d$, where $\Delta_d$ is the energy to break the {\it local} intrachain $d$-wave pair.
	%
	%
	%
	%
	When $t_\perp$ becomes larger ($t_\perp \gtrsim 0.25$), electrons hop between chains not as a pair but individually, so that the coherent pair hopping is suppressed.
	This is why only $\chi_d /\bar{\chi}_d$ saturates for large $t_\perp$, whereas $\chi_\rs /\bar{\chi}_\rs$ keeps increasing due to the single-particle process.
	From this viewpoint, both intra- and inter-chain Cooper pair forms impartially for $t_\perp \gtrsim 0.25$.
	Since both the intrachain (1D) spin-fluctuation and the interchain Josephson coupling help to form pairs, $\chi_d$ have a maximum at around $t_\perp \sim 0.25$.
	%

\begin{figure}[tb]
\begin{center}
\begin{flushleft}
\hspace{33mm}
\footnotesize
(a) \hspace{38mm} (c)
\end{flushleft}
\includegraphics[width=4.25cm]{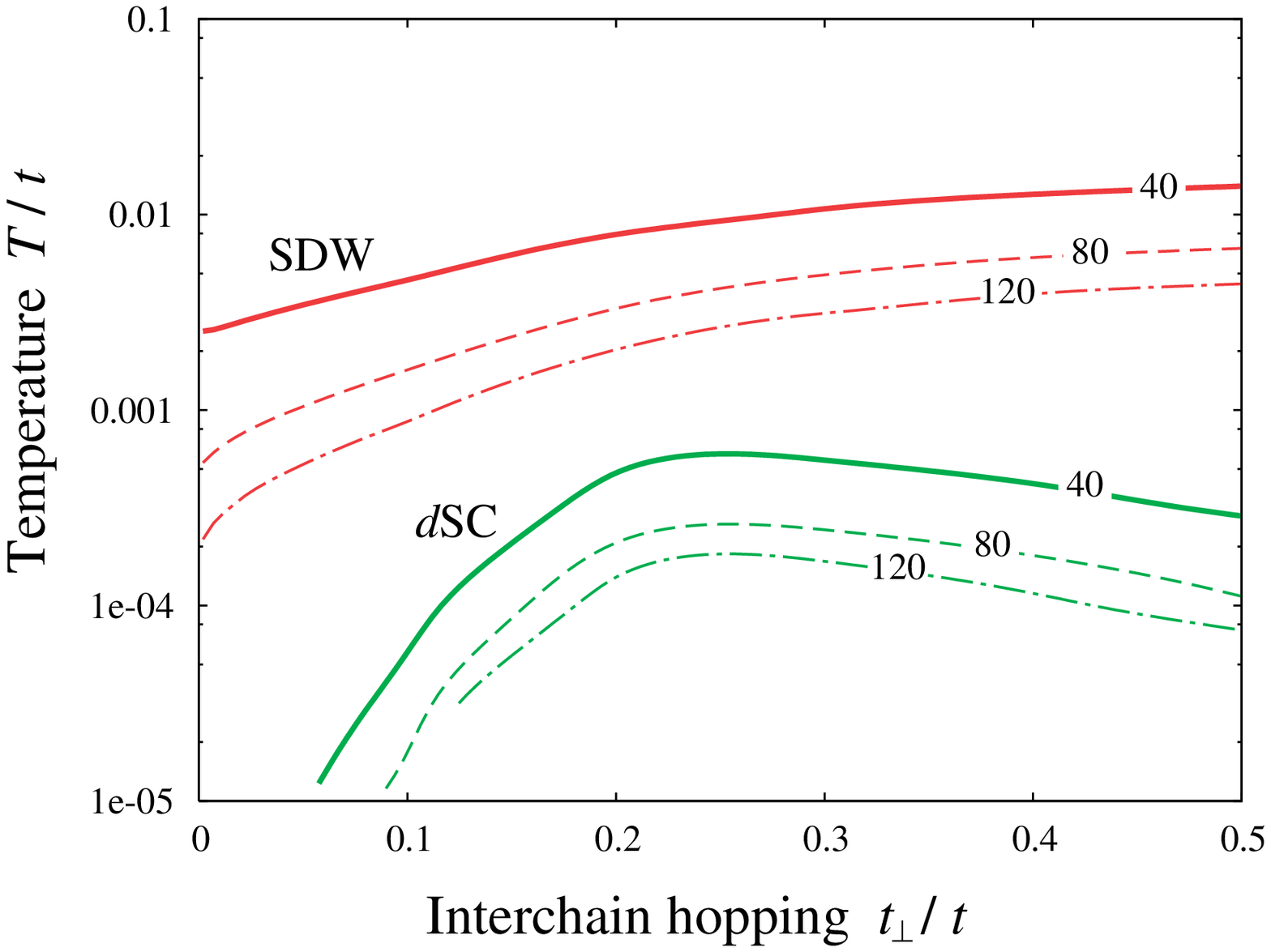}
\includegraphics[width=4.25cm]{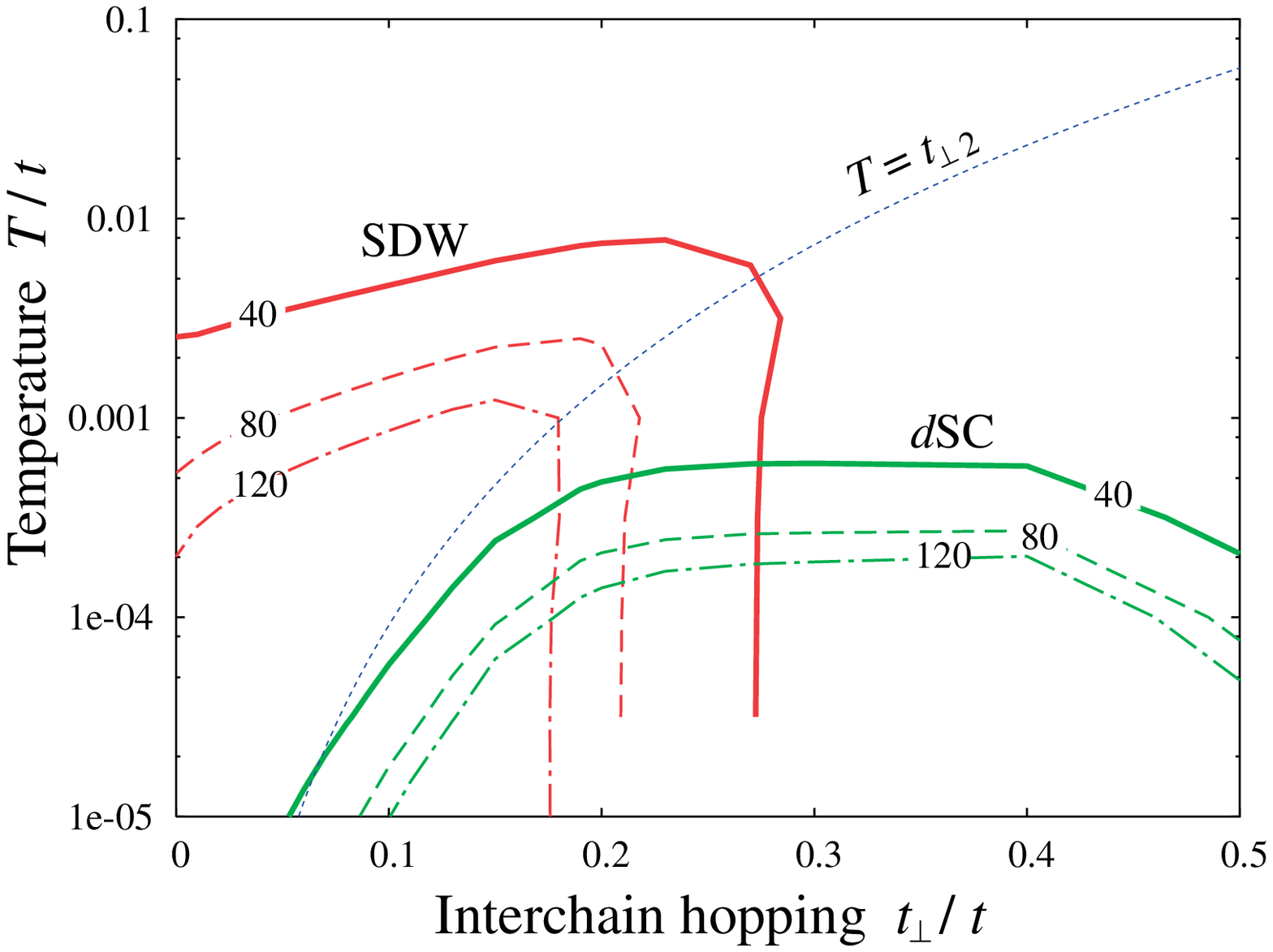}
\begin{flushleft}
\hspace{33mm}
\footnotesize
(b)
\end{flushleft}
\begin{center}
\includegraphics[width=7cm]{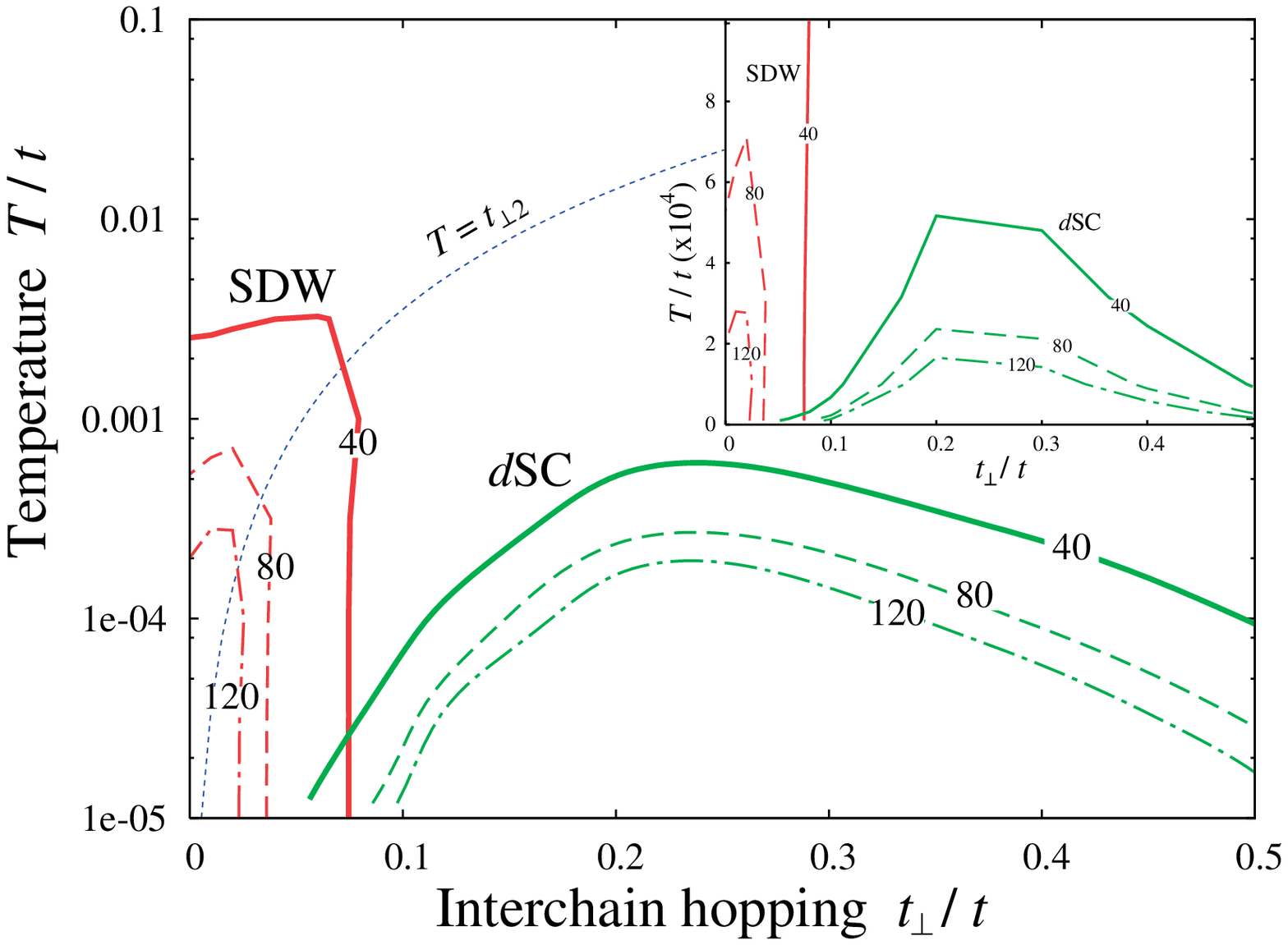}
\end{center}
\end{center}
\caption{
Contour plot of $\pi v_F \chi_\mu (T, t_\perp)$
with different nesting conditions: (a) $t_{\perp2}=0$, (b) $t_{\perp2}=-t_\perp^2 \cos k_F /4t \sin ^2 k_F$, (c) $t_{\perp2}/t\propto (t_\perp /t)^4$.
The insets of (b) is the linear plot of the low-temperature region.
}
\label{Phase}
\end{figure}

	%
	Next, we show the contour plot of $\pi v_F \chi_\mu (T, t_\perp)$ in Fig. \ref{Phase}.
	We plot three different situation of the nesting property, which correspond to different situation of the pressure effect.
	%
	%
	%
	%
	%
	%
	%
	%
	%
	%
	%
%
	%
	First, let us look into the role of the dimensionality exclusively by assuming $t_{\perp2}=0$ in Fig. \ref{Phase} (a).
	In this case, the Fermi surface is perfectly nested, and thus $\chi_\rs$ is always larger than $\chi_d$ and keeps increasing by $t_\perp$.
	Note that the contour lines of $\chi_d$ has a broad maximum and turns to decrease at around $t_\perp \sim 0.25 t$ even though the Fermi surface is perfectly nested and the spin-fluctuation is not weakened.
	This decrease is due to the individual interchain hopping of the electrons mentioned above\cite{SF}.
	%
	%
	
	%
	In the actual compounds, the pressure also has a role of increasing $t_{\perp2}$ and destroying the nesting property.
	%
	%
	For example, when we linearlize the tight-binding form for the anisotropic square lattice, $\xi (\bk) = -2t\cos k-2t_\perp \cos k_\perp -\mu$, we obtain $t_{\perp2}=-t_\perp^2 \cos k_F /4t \sin ^2 k_F$.
	We plot $\chi_\mu (T, t_\perp)$ with this condition in Fig. \ref{Phase} (b).
	The contour lines of $\chi_\rs$ suddenly fall off when they cross the line $T=t_{\perp2}$, since $\chi_\rs (T)$ is cut off at $T=t_{\perp2}$ as is shown in the inset of Fig. \ref{Fig1}.
	On the other hand, $\chi_d$ is not affected so much by the nesting property, so that the shape of its contour lines is similar to that for $t_{\perp2}=0$ except for the rather rapid decrease for $t_\perp \gtrsim 0.25t$.
	This indicates that $d$SC is insensitive to the nesting property (or the low-energy property of the spin fluctuation).\cite{Shimahara,DB,Fuseya2005b}
	As a result, the {\it superconducting dome} emerges.
	In such a situation, the superconducting correlation keeps increasing by pressure even when SDW correlation is suppressed, as is observed in (TMTTF)$_2$SbF$_6$.
	The superconducting correlation also has a maximum at around $t_\perp \sim 0.25t$.
	The rather rapid decrease of $\chi_d$ for $t_\perp \gtrsim 0.25t$ is due to the strongly reduced spin-fluctuations besides the individual interchain hopping.
	%
	
	%
	Then, we show another case, $t_{\perp2}/t=109.6 \times (t_\perp/t)^4$ in Fig. \ref{Phase} (c), where the Fermi surface keeps good nesting condition.
	%
	The behavior of SDW changes, but the essence does not change, namely, the contour lines fall down below $T=t_{\perp2}$.
	The behavior of $d$SC is not affected so much.

	These contour plots of $\chi_\mu (T, t_\perp)$ can be roughly considered as $P$-$T$ phase diagrams by introducing a small interplane coupling, $V_\mu$.
	In this case when $\chi_\mu (T, t_\perp)$ reaches the value $V_\mu^{-1}$, a phase transition occurs, so that Fig. \ref{Phase} represents to the phase diagram.
	We assume that the three dimensional transition temperatures are determined by the condition $1-V_\mu \chi_\mu (T, t_\perp)=0$ with a constant $V_\mu$.
	%
	%
	Note that, in order to estimate the real $\Tc$ and obtain the real $P$-$T$ phase diagram, we should take into account the details of the compounds and the pressure effect,
	such as the renormalization of the interplane process to $V_\mu$, and the pressure effects on $U/t$ and $V_\mu$.
	But the important statement here is that $\chi_d$ ($\Tc$) essentially increases by $t_\perp$ for small $t_\perp $, which have not been obtained by a conventional superconducting mechanism for higher dimensions, even with a simple assumption of the constant $V_\mu$.
	%
	%
	%
	Whether $\Tc$ keep increasing after the SDW phase is suppressed is determined only by the property of $\chi_\rs$, and hence by the energy scale of the nesting deviation $T=t_{\perp2}$.
	Therefore, even with a small diffrence of the nesting conditions, the qualitative property of the phase diagram can change.
	%
	%
	When the nesting is easily broken by the pressure, the phase diagram is like Fig. \ref{Phase} (b).
	The results of (TMTTF)$_2$SbF$_6$ corresponds to this situation.
	When the nesting property is solid against the pressure, the phase diagram becomes like Fig. \ref{Phase} (c).
	The TMTSF-salts would be in this situation.
	%
	%
	
	One may notice that the situation of Fig. \ref{Phase} (b) resembles that of high-$\Tc$ superconductors.
	Actually, the effects of the quantum interference between the Cooper and Peierls channels, which leads to the increase of $\Tc$ in the present case, are also discussed in the context of the origin of the pseudo-gap behavior of high-$\Tc$ superconductors\cite{Yanase}.
	But for the isotropic square-lattice, the RG method yields the monotonically decrease of $\Tc$\cite{Zanchi,Halboth,Honerkamp}.
	Therefore, in order to understand the $\Tc$ behavior of high-$\Tc$ superconductors, it would be necessary to take account of the strong correlation in the vicinity of the Mott insulator.\cite{Zhang,Ogata,Yokoyama}
	The present results of $\Tc$ is not relevant to the high-$\Tc$ superconductors;
	it is, instead, a characteristic behavior of Q1D systems.
	We also note that the increase of $\Tc$ is also realized when the long-range Coulomb interaction is finite and the symmetry of the superconductivity is $f$-wave triplet\cite{Fuseya2005b}.
	%

	
	In conclusion, we have calculated the response functions both for SDW, $\chi_\rs$, and $d$-wave singlet superconductivity ($d$SC), $\chi_d$, by the $N$-chain RG technique, and investigated the effect of dimensionality upon the spin- and superconducting-fluctuation.
	It is shown that, for the perfect nesting case, not only $\chi_\rs$ but also $\chi_d$ are enhanced by $t_\perp$.
	This increase of $\chi_d$ is realized only in the system where following two conditions are satisfied; i) the interference between Cooper and Peierls channels in higher energy region, and ii) $k_\perp$-dependence in couplings in lower energy region.
	These conditions are satisfied only in Q1D systems, and not in two (square) or three dimensional systems.
	%
	%
	The system gains largest superconducting correlations at around $t_\perp \simeq 0.25$, where both 1D fluctuation and spin fluctuation contribute to superconductivity.
	%
	%
	%
	%
	When the Fermi surface deviates from perfect nesting, $\chi_\rs (T)$ is cut off at $T=t_{\perp2}$, whereas $\chi_d (T)$ is not affected so much. 
	Consequently, $\chi_d$ increases even after the spin fluctuation is suppressed, which leads to the fact that $\Tc$ keeps increasing away from the SDW phase.
	This is also valid for the $f$-wave triplet pairings.
	%
	%
	Our results give a possible mechanism of the increasing-$\Tc$ behavior observed in (TMTTF)$_2$SbF$_6$.

	Fruitful discussions with M. Itoi, M. Hedo, Y. Uwatoko, Y. Suzumura, M. Tsuchiizu and K. Miyake are acknowledged.
	The present work was financially supported by Grants-in-Aid for Scientific Research on Priority Areas of Molecular Conductors (No. 15073210) from the Ministry of Education, Culture, Sports, Science and Technology, Japan, and Next Generation Supercomputing Project, Nanoscience Program, MEXT,Japan.
	Y. F. is supported by JSPS Research Fellowships for Young Scientists.


\end{document}